\documentclass[prl,aps,twocolumn,superscriptaddress,nofootinbib,longbibliography]{revtex4-1}
\usepackage{graphicx} 
\usepackage{dcolumn}  
\usepackage{bm}       
\usepackage{amsmath}
\usepackage{epsfig}
\usepackage{bbm}
\usepackage{color}
\usepackage{float}

\begin{document}
\title{Nanoscale transfer of angular momentum mediated by the Casimir torque}
\author{Stephen Sanders}
\affiliation{Department of Physics and Astronomy, University of New Mexico, Albuquerque, New Mexico 87131, United States}
\author{Wilton J. M. Kort-Kamp}
\affiliation{Theoretical Division, MS B262, Los Alamos National Laboratory, Los Alamos, New Mexico 87545, United States}
\author{Diego A. R. Dalvit}
\affiliation{Theoretical Division, MS B213, Los Alamos National Laboratory, Los Alamos, New Mexico 87545, United States}
\author{Alejandro Manjavacas}
\email[Corresponding author: ]{manjavacas@unm.edu}
\affiliation{Department of Physics and Astronomy, University of New Mexico, Albuquerque, New Mexico 87131, United States}

\date{\today}

\begin{abstract}
Casimir interactions play an important role in the dynamics of nanoscale objects. Here, we investigate the noncontact transfer of angular momentum at the nanoscale through the analysis of the Casimir torque acting on a chain of rotating nanoparticles. We show that this interaction, which arises from the vacuum and thermal fluctuations of the electromagnetic field, enables an efficient transfer of angular momentum between the elements of the chain. Working within the framework of fluctuational electrodynamics, we derive analytical expressions for the Casimir torque acting on each nanoparticle in the chain, which we use to study the synchronization of chains with different geometries and to predict unexpected dynamics, including a \textit{rattleback}-like behavior. Our results provide new insights into the Casimir torque and how it can be exploited to achieve efficient noncontact transfer of angular momentum at the nanoscale, and therefore have important implications for the control and manipulation of nanomechanical devices.
\end{abstract}
\maketitle


\section{Introduction}

It is well established that light carries angular momentum \cite{AB12}. The simplest evidence of this phenomenon is the possibility of spinning nanostructures by illuminating them with circularly polarized light \cite{B1936,FNH98}.  The angular momentum of light is also manifested in Casimir interactions \cite{DMR11} that originate from the vacuum and thermal fluctuations of the electromagnetic field \cite{M94,WDT16}. For instance, two parallel plates made of birefringent materials with in-plane optical anisotropy have been shown to experience a Casimir torque that rotates them to a configuration in which the two optical axes are aligned \cite{PW1972,B1978,MIB05,SM17,SGP18}. A similar phenomenon is predicted to occur for two nanostructured plates \cite{GGL15}, or for a nanorod placed above a birefringent plate \cite{XL17}.

Vacuum and thermal fluctuations also produce friction on rotating nanostructures; for instance, it has been predicted that a nanoparticle rotating in vacuum experiences a Casimir torque that slows its angular velocity and eventually stops it \cite{ama7,ama9}. A similar effect is also expected for a rotating pair of atoms \cite{BL15}. The Casimir torque can be enhanced by placing the particle near a substrate \cite{ama19,DK12}, which also leads to a lateral Casimir force \cite{ama50,QW19}, or by using materials with magneto-optical response \cite{PXG17}. 
The origin of this torque can be found in the imbalance of the absorption and emission of left- and right-handed photons caused by the rotation of the particle. 
Although Casimir interactions produce friction and stiction that affect the dynamics of nanomechanical devices \cite{MC10}, they also offer new opportunities for the noncontact transfer of momentum; for example, if a rotating nanostructure is placed near another structure, the Casimir torque that slows the first one down will necessarily accelerate the other one \cite{MJK12,casimir_torque,LS16,V17,AE17,RMP17}.

In this paper, we investigate the noncontact transfer of angular momentum at the nanoscale by analyzing the Casimir torque acting on a chain of rotating nanoparticles, which contains an arbitrary number of elements. This system can be thought of as a Casimir analog to a chain of particles rotating in a viscous medium, whose motion is coupled through a drag force.  We obtain analytical expressions describing the Casimir torque and exploit them to study the free rotational dynamics of chains with different geometries.
We show that the synchronization of the angular velocities of the particles can happen in timescales as short as seconds for realistic structures. Furthermore, we predict exotic behaviors in chains with inhomogeneous particle sizes and separations, including {\it rattleback}-like dynamics \cite{GH1988,KN17}, in which the sense of rotation of a particle changes several times before synchronization, as well as configurations for which angular momentum is never transferred to a selected particle in the chain.
We analyze, as well, the steady-state distribution of angular velocities in chains in which one or multiple particles are externally driven at a constant angular velocity.
The results of this work show that the Casimir torque provides a new mechanism for the efficient transfer of angular momentum between nanoscale objects without requiring them to be in contact.

\section{Results}

\begin{figure}
\begin{center}
\includegraphics[width=85mm,angle=0,clip]{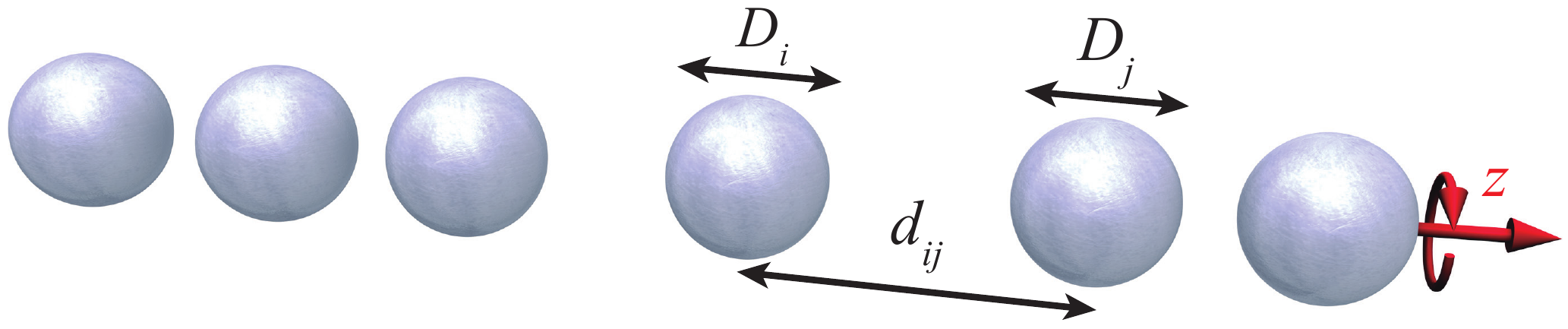}
\caption{System schematics. The system under study consists of a chain of $N$ nanoparticles with diameters $D_i$ separated by a center-to-center distance $d_{ij}$. The nanoparticles are assumed to rotate around the axis of the chain.}  \label{fig1}
\end{center}
\end{figure}

The system under study is depicted in Fig.~\ref{fig1}. It consists of a chain of $N$ spherical nanoparticles with diameters $D_i$ separated by  center-to-center distances $d_{ij}$. Each nanoparticle can rotate with an angular velocity $\Omega_i$ around the axis of the chain, which we choose as the $z$ axis. We assume that the size of the particles is much smaller than the relevant wavelengths of the problem, which are determined by the temperature, the material properties, and the angular velocities of the particles, and consider geometries obeying $d_{ij}\geq \frac{3}{2}\max(D_i,D_j)$. This allows us to model each particle as a point electric dipole with a frequency-dependent polarizability $\alpha_i(\omega)$.
Within this approximation, the Casimir torque acting on particle $i$ can be written as  $M_{i} = \langle\mathbf{p}_i\times\mathbf{E}_i \rangle\cdot \hat{\mathbf{z}}$ \cite{J99}, where the brackets  stand for the average over vacuum and thermal fluctuations, while $\mathbf{p}_i$ and $\mathbf{E}_i$ are, respectively, the self-consistent dipole and field at particle $i$, which originate from: (i) the fluctuations of the dipole moment of each particle $\mathbf{p}_j^{\rm{fl}}$ and (ii) the fluctuations of the field $\mathbf{E}_j^{\rm{fl}}$. 
Then, solving $\mathbf{p}_i$ and $\mathbf{E}_i$ in terms of  $\mathbf{p}_j^{\rm{fl}}$ and $\mathbf{E}_j^{\rm{fl}}$ and taking the average over fluctuations using the fluctuation-dissipation theorem \cite{N1928,CW1951,NH06}, one can write the Casimir torque as $M_i=M_i^+ - M_i^-$, where $M_i^{\pm}$ is given by (see the Appendix  for the detailed derivation) 
\begin{equation}
M_i^{\pm} = -\frac{2\hbar}{\pi}\int_0^{\infty}\!\!\! d\omega \sum^N_{j=1}\left[ \Gamma_j^{0\pm} \delta_{ij} + \Gamma^{\pm}_{ij} \right] \left[n_j(\omega_{j}^{\mp})-n_0(\omega)\right].\nonumber
\end{equation}
Here, $\omega_{i}^{\pm}=\omega\pm\Omega_i$ and $n_{i}(\omega)=[\exp(\hbar \omega/k_B T_i)-1]^{-1}$ is the Bose-Einstein distribution at temperature $T_i$, with $T_0$ being the  temperature of the surrounding environment. 
Furthermore, $\Gamma^{0\pm}_{i}=\frac{2\omega^3}{3c^3}{\rm Re}\left\{2A^{\pm}_{ii}-1\right\} \chi^{\pm}_i(\omega)$ corresponds to the contribution to the Casimir torque produced by the environment, while 
\begin{equation}
\Gamma_{ij}^{\pm}=\delta_{ij} {\rm Im}\left\{S^{\pm}_{ii} \right\} \chi^{\pm}_i(\omega)- \left|S^{\pm}_{ij}\right|^2\chi^{\pm}_i(\omega)\chi^{\pm}_j(\omega)\nonumber
\end{equation}
is the contribution arising from the particle-particle interaction. Physically, the first of these contributions corresponds to the exchange of angular momentum between each particle and the environment, while the second one is associated with the exchange of angular momentum between the particles.
In these expressions, $A^{\pm}_{ij}$ are the components of the $N\times N$ matrix  $\mathbf{A}^{\pm}=\left[\mathcal{I}-\bm{\alpha}^{\pm}\mathbf{G}\right]^{-1}$, where $\bm{\alpha}^{\pm}$ is a diagonal matrix whose components are the effective polarizabilities of the particles $\alpha^{\pm}_{{\rm eff},i}(\omega)$ as seen from the frame at rest, $\mathbf{G}$ is the dipole-dipole interaction matrix with components $G_{ij} = (1-\delta_{ij})\exp(ikd_{ij})[(kd_{ij})^2+ikd_{ij}-1]/d^3_{ij}$, and $k=\omega/c$ is the wave number. Furthermore,  $\chi^{\pm}_i(\omega)={\rm Im}\left\{\alpha^{\pm}_{{\rm eff},i}(\omega)\right\}-\frac{2}{3}k^3\left|\alpha^{\pm}_{{\rm eff},i}(\omega)\right|^2$ and $S^{\pm}_{ij}=\sum^N_{m=1} A_{mi}^{\pm}G_{mj}$  (see the Appendix). We want to remark that our results include radiative corrections, which are crucial to correctly describe the contributions to the torque produced by the environment \cite{N15}. It is also important to notice that the Casimir torque described here is a dissipative effect, and therefore it depends strongly on the temperature of the particles and the environment.

The calculation of the effective polarizability $\alpha^{\pm}_{{\rm eff},i}(\omega)$ is subtle. In the past, this quantity has been computed assuming that the intrinsic response of the system was independent of the rotation, which resulted in an effective polarizability $\alpha^{\pm}_{{\rm eff},i}(\omega)=\alpha_i(\omega_i^{\mp})$ that only accounted for the Doppler-shift produced by the rotation \cite{ama7,ama9,ama19,ama50,MGK13,MGK13_2,MJK14,LS16}. However, recently \cite{PXG17,PXG19}, it has been pointed out that the inclusion of the Coriolis and centrifugal effects gives rise to corrections that, for the case of spherical particles, cancel the effect of the Doppler-shift and introduce a dependence on $\Omega^2$. As a consequence of this, the effective polarizability of a rotating sphere becomes $\alpha^{\pm}_{{\rm eff},i}(\omega)=\alpha_i(\omega)+\mathcal{O}(\Omega^2)$, as shown in the Appendix. Notice that the component of the polarizability along the rotation axis, which does not contribute to the Casimir torque, is not affected by the rotation.

Under realistic conditions, $\Omega_i$ is smaller than both the thermal frequency $k_BT/\hbar$ ($\approx6\,$THz at room temperature) and the frequencies of the optical modes of the nanoparticles ($\approx 28\,$THz for SiC). This allows us to expand $M_i^{\pm}$ in powers of $\Omega_i$ and retain the lowest nonvanishing order, which is the linear one. This linear approximation is accurate for angular velocities up to $\Omega_i/2\pi\sim 100\,$GHz, as shown in Fig.~\ref{figS1}.  Importantly, within this limit, the corrections to the effective polarizability discussed above become irrelevant.
Assuming all particles and the environment have the same finite temperature, we can write the following equation for their angular velocities 
\begin{equation}
\dot{\Omega}_i = \sum_{j=1}^NH_{ij}\Omega_j, \nonumber
\end{equation}
where the matrix $\mathbf{H}$ has components  $H_{ij}=h_i^0\delta_{ij}+h_{ij}$ with
\begin{equation}
 \begin{pmatrix}h^0_i \\ h_{ij}
 \end{pmatrix}= \frac{4\hbar}{\pi I_i}\int_0^{\infty}\!\!\! d\omega 
  \begin{pmatrix}\Gamma^{0}_i \\ \Gamma_{ij}
  \end{pmatrix}
 \frac{\partial n(\omega)}{\partial \omega}. \nonumber
\end{equation} 
 Here,  $I_i = \pi\rho D_i^5/60$ is the moment of inertia of the nanoparticles, and $\rho$ their mass density. The expressions for $\Gamma_{i}^0$ and $\Gamma_{ij}$ are obtained from the corresponding definitions above by setting all angular velocities to zero.

\begin{figure}
\begin{center}
\includegraphics[width=85mm,angle=0,clip]{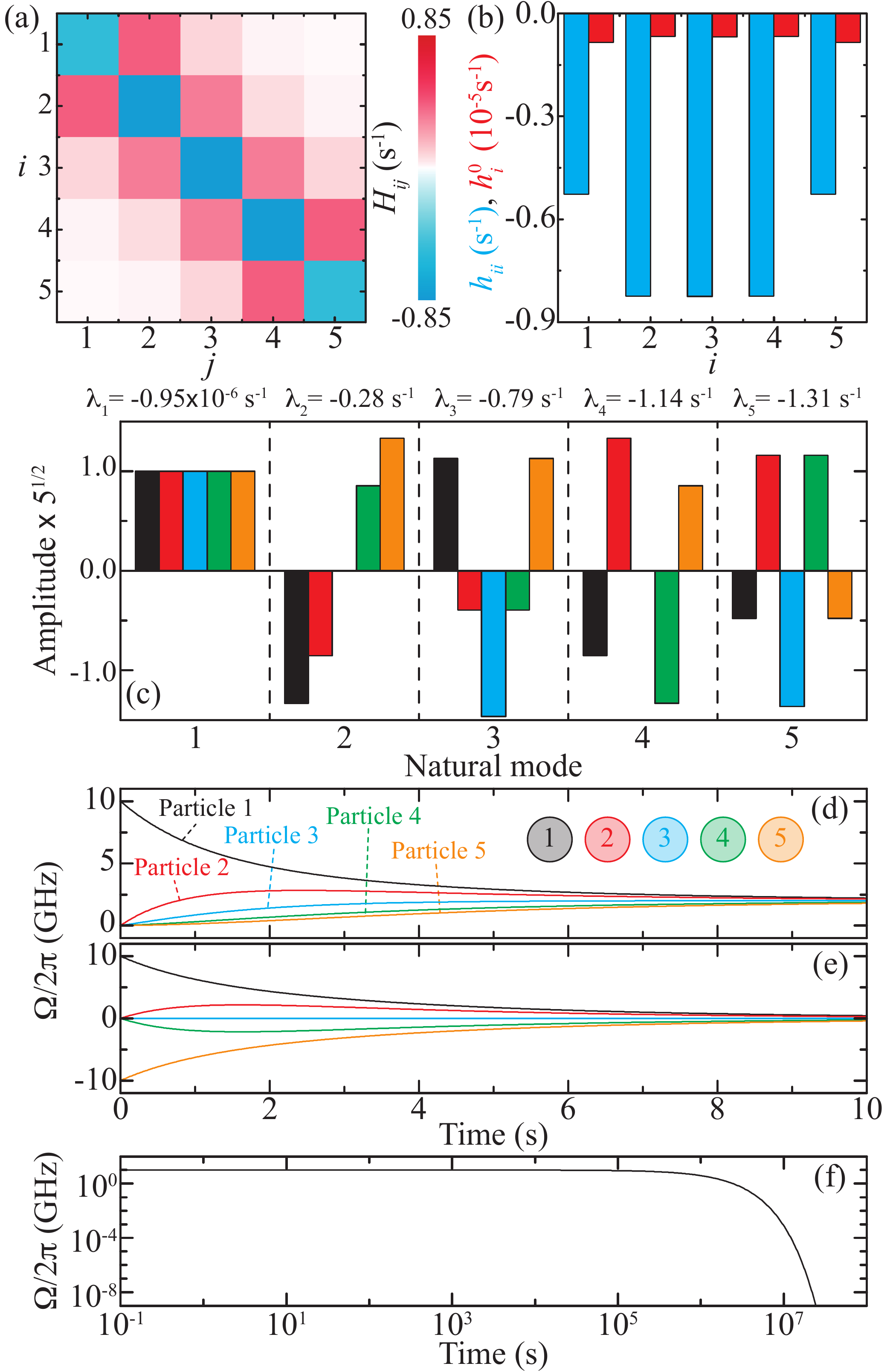}
\caption{Rotational dynamics. (a) $H_{ij}$ for a chain of $N=5$ identical particles with $D=10\,$nm made of SiC, which are uniformly distributed with a center-to-center distance $d=1.5D$. (b) Diagonal components of $h_{ij}$ (blue) and the corresponding $h^0_{i}$ (red). Notice the different scale. (c) Natural modes of the chain obtained by diagonalizing $H_{ij}$ and the corresponding decay rates. (d)-(f) Temporal evolution of the angular velocity of each particle in the chain for three different initial conditions.}  \label{fig2}
\end{center}
\end{figure}

We can hence study the rotation dynamics of a chain by analyzing its natural decay rates and modes given, respectively, by the eigenvalues and eigenvectors of $\mathbf{H}$. As an initial example, we analyze a chain with $N=5$ SiC spheres, all of them with identical diameter $D=10\,$nm, that are uniformly distributed with a center-to-center distance $d=1.5D$. Throughout this work, unless stated otherwise, we assume that all of the particles remain at the same temperature as the environment, and set that temperature to $300\,$K. This assumption is a good approximation for the systems under consideration, as discussed in the Appendix and  Fig.~\ref{figS2}.
The polarizability of the particles is obtained from the dipolar Mie coefficient \cite{paper112} with the dielectric function of SiC modeled as $\varepsilon(\omega)=\varepsilon_{\infty}\left[1+(\omega^2_{\rm L}-\omega^2_{\rm T})/(\omega^2_{\rm T}-\omega^2-i\omega\gamma)\right]$, with $\varepsilon_{\infty}=6.7$, $\hbar\omega_{\rm T}=98.3\,$meV,  $\hbar\omega_{\rm L}=120\,$meV, and $\hbar\gamma=0.59\,$meV \cite{P1985}. With this choice of size and material, the optical response of the nanoparticles is dominated by a phonon polariton at $\approx28\,$THz.
Fig.~\ref{fig2}(a) shows the different components of $\mathbf{H}$ for the chain under analysis. 
As a consequence of the prevalent role of the near-field coupling, the matrix is almost tridiagonal. Furthermore, the diagonal elements are all negative, with the contribution of the environment, $h_i^0$, being five orders of magnitude smaller than that of the particle-particle interaction $h_{ij}$, as shown in panel (b). Interestingly, $\mathbf{H}$ is a negative definite matrix (\textit{i.e.}, all of its eigenvalues are strictly negative), which ensures that, in absence of external driving, the angular velocities decay to zero at large times. 

The natural modes of the chain, together with the corresponding decay rates, are displayed in Fig.~\ref{fig2}(c). The first natural mode corresponds to all particles rotating with the same angular velocity, which makes the contribution arising from the particle-particle interaction vanish, leaving only the Casimir torque produced by the environment. This results in a decay rate $\lambda_1=-0.95\times10^{-6}\,$s$^{-1}$ (in perfect agreement with the stopping time calculated in \cite{ama7} for single particles), much smaller than those of the remaining modes, which are all on the order of s$^{-1}$. In these other modes, the particles rotate with angular velocities of different magnitude and sign, but, as expected from the symmetry of the system, the modes are either even or odd with respect to the central particle, which, consequently, is at rest in the odd modes.

\begin{figure}
\begin{center}
\includegraphics[width=85mm,angle=0,clip]{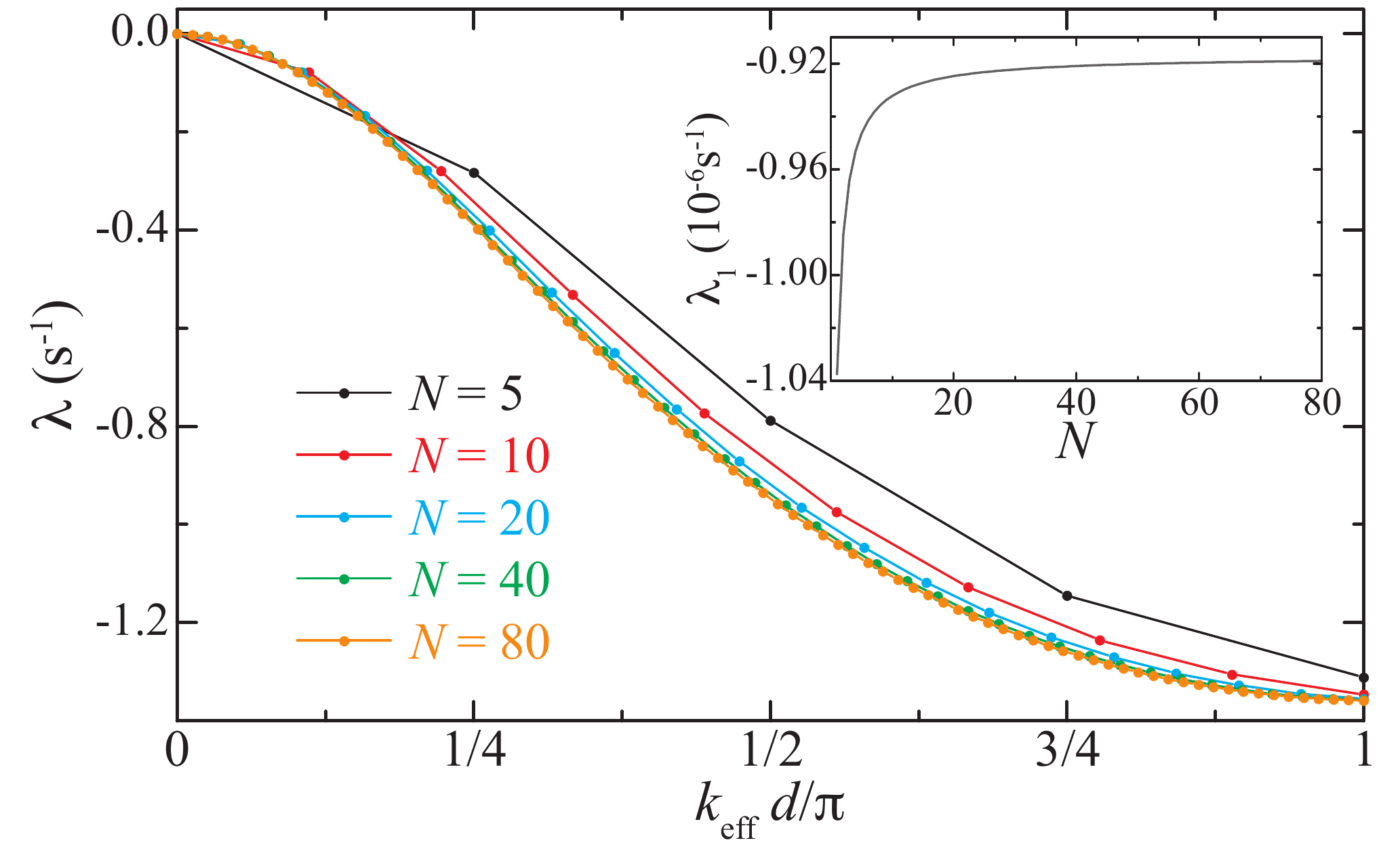}
\caption{ Natural decay rates. Evolution of the natural decay rates for chains with different $N$ (as indicated in the legend) plotted as a function of the effective momentum $k_{\rm eff}$.  The inset shows the evolution of the slowest decay rate $\lambda_1$ as a function of $N$. In all of the cases, $D=10\,$nm and $d=1.5D$. }  \label{fig3}
\end{center}
\end{figure}

\begin{figure*}
\begin{center}
\includegraphics[width=170mm,angle=0,clip]{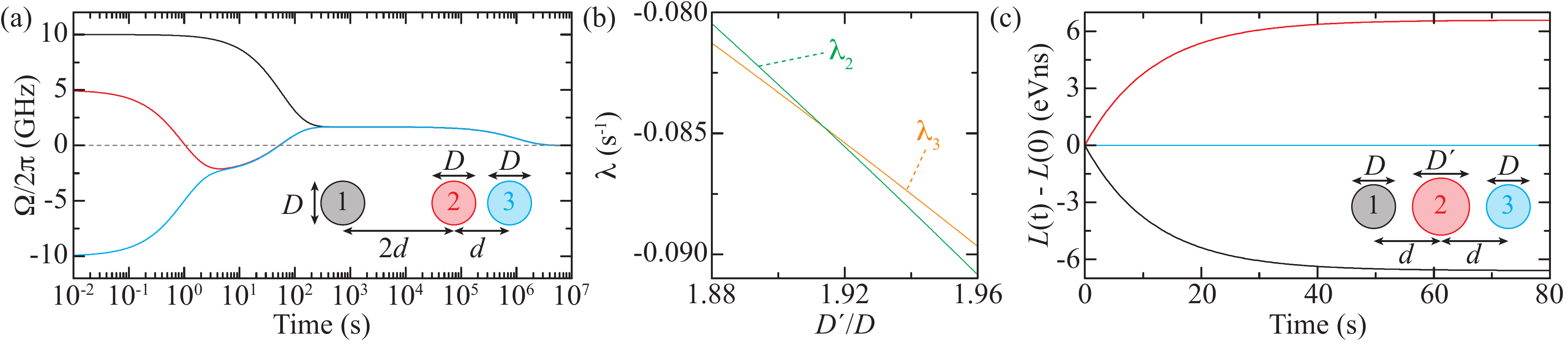}
\caption{Exotic rotational dynamics. (a) Temporal evolution of the angular velocity of the particles of a chain with $N=3$, arranged as shown in the inset, where $D=10\,$nm and $d=1.5D$. (b) Crossing of the second and third decay rates  (\textit{i.e.}, $\lambda_2$ and $\lambda_3$) of the $N=3$ chain shown in the inset of panel (c) when varying the ratio $D'/D$. In this case, $D=10\,$nm and $d=3D$. (c) Temporal evolution of the angular momentum, $L(t)=I\Omega(t)$, for each of the three particles of the chain depicted in the inset, assuming $D'/D=1.914$. We analyze two different cases with initial angular velocities given by $\Omega_1/2\pi=10\,{\rm GHz}+\Omega_c/2\pi$, $\Omega_2/2\pi=-0.39\,{\rm GHz}+\Omega_c/2\pi$, and $\Omega_3=\Omega_c$. In one case, $\Omega_c=0$ while, in the other, $\Omega_c/2\pi=5\,$GHz. However, since there is no transfer of angular momentum to particle $3$, both cases produce exactly the same results.
} \label{fig4}
\end{center}
\end{figure*}

Figures~\ref{fig2}(d)-(f) show the temporal evolution of the angular velocities of the particles for three different initial conditions. Specifically, in panel (d), only particle $1$ (black) is initially rotating with $\Omega_1/2\pi=10\,$GHz, an angular velocity that is within experimental reach, as recently demonstrated \cite{RDH18,AXB18}.
As time evolves, the rotation is transferred  through the chain to the other particles, resulting in a synchronized rotation dynamics after a few seconds, in which the initial angular momentum is uniformly distributed among all particles. 
The situation is different when particle $5$ (yellow) is also set to rotate initially at $\Omega_5=-\Omega_1$. In this case, since the initial angular momentum of the system is zero, the rotation of the particles ends after a few seconds, as shown in panel (e). It is important to notice that any initial state with finite angular momentum must have a nonzero overlap with the first natural mode. In these cases, after synchronization, the angular velocity of the particles gradually decreases and eventually stops on a much larger timescale $\approx |\lambda_1|^{-1}$, as a consequence of the Casimir torque produced by  the environment (\textit{i.e.}, $h^0_i$). This can be seen in panel (f), where we plot the temporal evolution of the chain when initialized in the first natural mode.

The rotation dynamics of chains with arbitrarily large $N$ can be understood in a similar way by analyzing the corresponding decay rates and natural modes. 
In Fig.~\ref{fig3}, we plot the decay rates of chains with different $N$ as a function of an effective momentum $k_{\rm eff}$, defined as $k_{\rm eff}d/\pi=(n-1)/(N-1)$, where $n$ is the index labeling the natural modes. The effective momentum is directly proportional to the number of nodes of the natural mode, \textit{i.e.}, the number of times that its components change sign. Examining Fig.~\ref{fig3}, we observe that, as $N$ increases, the decay rates converge to a curve that resembles the dispersion relation of the transversal mode of the infinite chain \cite{WF04}. As in the $N=5$ case analyzed before, the first mode always corresponds to all particles rotating with the same angular velocity (see Fig.~\ref{figS3}), which means that only the environment contributes to the Casimir torque, thus resulting in a decay rate $\lambda_1$ with values $\approx 10^{-6}\,$s$^{-1}$ for any $N$, as shown in the inset. 
As $k_{\rm eff}$ increases, the corresponding modes show a more complicated pattern in which neighboring particles rotate with increasingly different angular velocities (see Fig.~\ref{figS3}). This makes the contribution arising from the particle-particle interaction increase, leading to much faster decay rates.

We can use the insight obtained from the analysis of the natural modes and decay rates to study the behavior of chains with more exotic rotation dynamics. In particular, 
by breaking the uniformity in the particle separation, it is possible to obtain a \textit{rattleback}-like rotation dynamics \cite{GH1988,KN17} in which a particle reverses its sense of rotation multiple times during the synchronization process. This is shown in Fig.~\ref{fig4}(a) for a chain of $N=3$ SiC particles with $D=10\,$nm, in which the central particle is, respectively, at distance $2d$ and $d$ (with $d=1.5D$) from the left and right particles, as depicted in the inset. Analyzing the dynamics of the system, we observe that particle $2$ (red) changes its sense of rotation two times before all particles synchronize. This happens because this particle synchronizes first with particle $3$, due to the smaller distance that separates them, which results in a larger coupling, and, therefore, Casimir torque, among them. After that, both particle $2$ and $3$ have to synchronize with particle $1$ and, since the total initial angular momentum is positive, the synchronized angular velocities have to be positive. Interestingly, it is possible to obtain more reversals in the sense of rotation using chains with larger $N$, as shown in Fig.~\ref{figS4} for a chain with five particles. In all cases, after synchronization, the rotation velocities of the particles decay as a consequence of the Casimir torque produced by  the environment. 

Another interesting situation arises when the particles in the chain have different sizes. In particular, it is possible to make the decay rates of two different modes equal, as shown in Fig.~\ref{fig4}(b) for the $N=3$ chain depicted in the inset of panel (c), with $D=10\,$nm and $d=3D$. Clearly, as the diameter of the central particle $D'$ increases, the decay rates of the second (green curve) and third (yellow curve) natural modes cross each other. At the crossing point $D'=1.914D$, the degeneracy of the natural modes allows us to prepare the system in an initial state such that angular momentum is only transferred between the central and one of the side particles, without altering the dynamics of the remaining one.
We analyze two different examples of this behavior in panel (c), where we plot the temporal evolution of the angular momentum, $L(t)=I\Omega(t)$, for each of the three particles. The corresponding initial angular velocities are: $\Omega_1/2\pi=10\,{\rm GHz}+\Omega_c/2\pi$, $\Omega_2/2\pi=-0.39\,{\rm GHz}+\Omega_c/2\pi$, and $\Omega_3=\Omega_c$  (see the Appendix). In the first example, we choose $\Omega_c=0$, whereas in the second one, $\Omega_c/2\pi=5\,$GHz. 
In both cases, as expected, the angular momentum of particles $1$ (black curve) and $2$ (red curve) changes identically, but with opposite signs, while the angular momentum of particle $3$ remains completely unchanged (blue curve).

\begin{figure}
\begin{center}
\includegraphics[width=85mm,angle=0,clip]{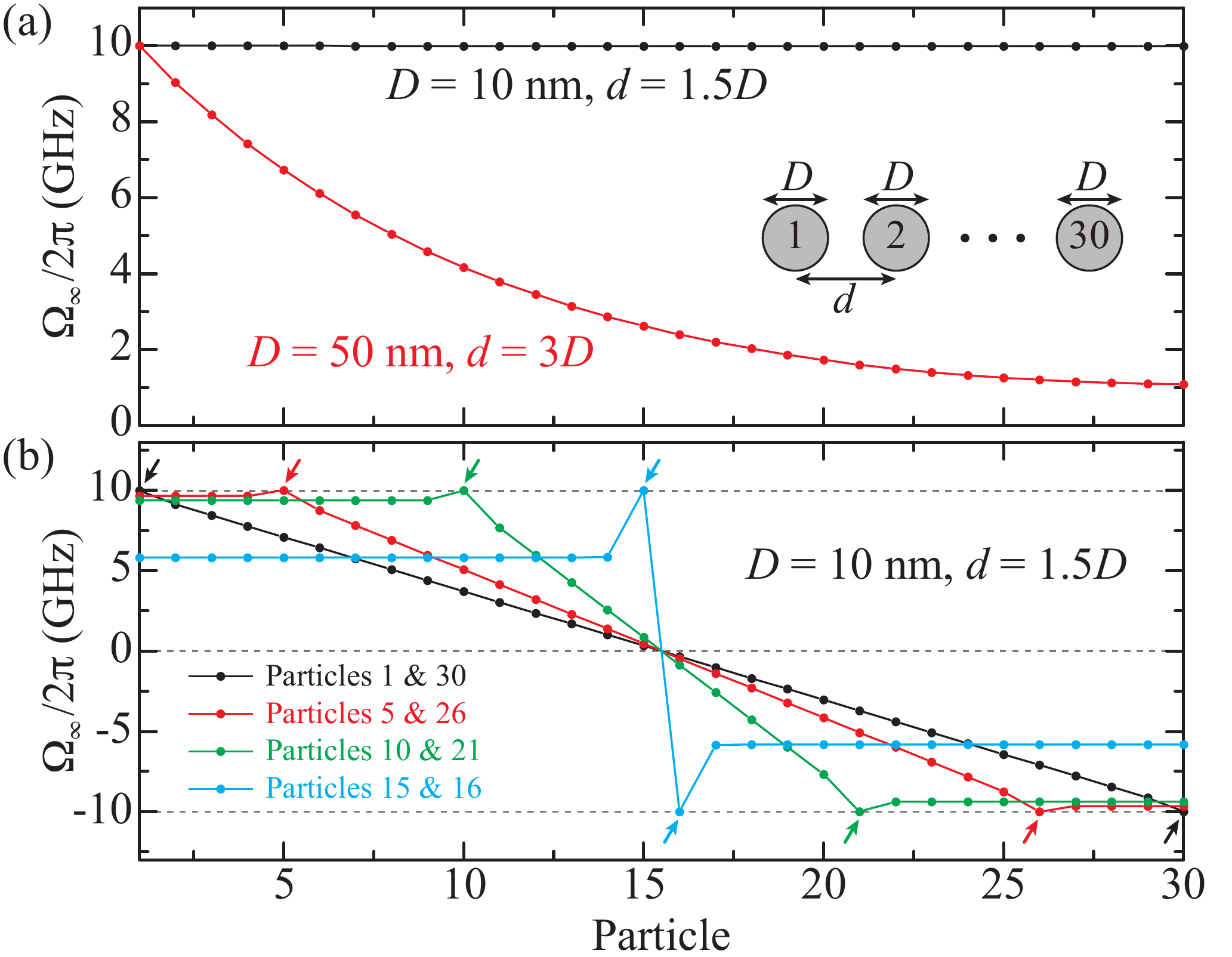}
\caption{Driven rotational dynamics. (a) Steady-state angular velocity for the particles of a $N=30$ chain, in which particle $1$ is externally driven at an angular velocity $10\,$GHz. All of the particles have the same diameter $D$ and are uniformly distributed with a center-to-center distance $d$, as shown in the inset. (b) Same as (a), but, in this case, two particles, which are indicated with small arrows (see legend), are externally driven at angular velocities $10\,$GHz and $-10\,$GHz, respectively. }  \label{fig5}
\end{center}
\end{figure}
  
So far, we have analyzed systems in free rotation, in which the initial angular momentum is transferred among the particles in the chain and is eventually dissipated into the surrounding environment. The situation is different when one or more particles in the chain are externally driven so their angular velocities remain constant. These particles act as a continuous source of angular momentum, which is transferred to the rest of the particles of the chain, and, as a consequence, after some transient evolution, the whole system reaches a steady-state rotation dynamics. As an initial example, we consider the chain of $N=30$ identical SiC nanoparticles shown in the inset of Fig.~\ref{fig5}(a), in which  particle $1$ is externally driven at a constant angular velocity $\Omega_1/2\pi = 10\,$GHz. We consider two different combinations of $D$ and $d$ for which we calculate the corresponding steady-state angular velocities, which are displayed in Fig.~\ref{fig5}(a). These velocities are determined by the interplay between the contributions to the Casimir torque arising from the environment and particle-particle interactions, 
with the former being the mechanism that dissipates angular momentum out of the chain, and the latter being the one that mediates its transfer between the particles. 
For $D=10\,$nm and $d=1.5D$ (black curve), the environment contribution is much smaller than that of the particle-particle interaction and, consequently, the angular velocities are almost identical. However, for $D=50\,$nm and $d=3D$ (red curve), the difference between the two contributions is reduced (see Fig.~\ref{figS5}), and, therefore, the angular velocities significantly decrease as we move away from the driven particle. This demonstrates that an efficient transfer of angular momentum is possible in systems for which $|h_{ii}|\gg |h_i^0|$.

Another interesting situation arises when two particles in the chain are driven at opposite angular velocities. In this case, as shown in Fig.~\ref{fig5}(b), the particles located between them have steady-state angular velocities with values uniformly distributed between those of the driven particles. On the other hand, the particles outside display almost constant velocities, which are determined by the separation between the driven particles.

\section{Discussion}

In summary, we have shown that the Casimir torque enables an efficient transfer of angular momentum at the nanoscale. To that end, we have analyzed the dynamics of chains of rotating nanoparticles with an arbitrary number of elements. We have found that this noncontact interaction leads to rotational dynamics happening on timescales as fast as seconds for nanoscale particles, which corresponds to torques on the order of  $10^{-27}\,$Nm for angular velocities of $10\,$GHz, both of which are within experimental reach \cite{XL17,MGV18,RDH18,AXB18,RTS18}.
We have derived an analytical formalism describing the rotational dynamics of these systems, which is based on the analysis of the natural modes of the chain,
 and exploited it to reveal unexpected behaviors. These include a \textit{rattleback}-like dynamics, in which the sense of rotation of a particle changes multiple times before synchronization, as well as configurations for which a selected particle is left out of the angular momentum transfer process. Furthermore, by analyzing the steady-state angular velocities of systems in which one or more particles are externally driven, we have established the conditions under which an efficient transfer of angular momentum can be achieved.
A possible experimental verification of our results would involve the use of nanoparticles trapped with optical tweezers \cite{MGV18,RDH18,AXB18}, or molecules with rotational degrees of freedom, such as  as chains of fullerenes \cite{WIZ08},  whose angular velocity can be detected through measurements of rotational frequency shifts \cite{BB97,MHS05}. The presented results describe a new mechanism for the noncontact transfer of angular momentum at the nanoscale, which brings new opportunities for the control of nanomechanical devices.

\section{Acknowledgments}
This work has been partially sponsored by the U.S. National Science Foundation (Grant ECCS-1710697).
We would like to thank the UNM Center for Advanced Research Computing, supported in part by the National Science Foundation, for providing the high performance computing resources used in this work. W. K. K. and D. D. acknowledge financial support from LANL LDRD program. We are also grateful to Prof. Javier Garc\'{i}a de Abajo for valuable and enjoyable discussions.

\onecolumngrid
\appendix
\section{Appendix}\label{ap}

\subsection{Derivation of Casimir Torque}

The system under consideration is depicted in Fig.~\ref{fig1}. It consists of a linear chain with $N$ spherical nanoparticles. Each of the particles has a diameter $D_i$ and is separated from its neighbors by a center-to-center distance $d_{ij}$. All of the particles are allowed to rotate with arbitrary angular velocity $\Omega_i$ around the axis of the chain, which we choose to be the $z$ axis. As discussed in the paper, we assume that the size of the particles is much smaller than the relevant wavelengths of the problem, which are determined by the temperature, the angular velocities, and the material properties of the particles, and use geometries obeying $d_{ij}\geq \frac{3}{2}\max(D_i,D_j)$. Within these limits, we can model each particle as a point electric dipole with a frequency-dependent polarizability $\alpha_i(\omega)$. This allows us to write the torque acting on particle $i$ as $M_{i} = \left\langle\mathbf{p}_i(t)\times \mathbf{E}_{i}(t)\right\rangle\cdot\mathbf{\hat{z}}$, where $\langle \rangle$ represents the average over fluctuations. Working in the frequency domain $\omega$, defined via the Fourier transform
$\mathbf{p}_i(t) = \int_{-\infty}^\infty \!\frac{d\omega}{2\pi}  \mathbf{p}_i(\omega)e^{-i\omega t}$,
for the dipole moment, and similarly for other quantities, and using the circular basis defined as: $\mathbf{\hat{e}}_{\pm} = (1/\sqrt{2})(\mathbf{\hat{x}}\pm i\mathbf{\hat{y}})$, and $\mathbf{\hat{e}}_0=\mathbf{\hat{z}}$, we can rewrite the torque as $M_{i} = M_{i}^+ - M_{i}^-$ with

\begin{equation}\label{Miz}
M_{i}^\pm = i\int_{-\infty}^\infty \! \frac{d\omega d\omega'}{4\pi^2} e^{-i(\omega-\omega')t} \langle {p_i^{\pm}}^\ast(\omega')E_i^\pm(\omega)\rangle.
\end{equation}
In this expression, $^\ast$ denotes the complex conjugate, while $p^{\pm}_{i}(\omega)$ and $E^{\pm}_{i}(\omega)$ are the self-consistent dipole moment and electric field in particle $i$, which originate from the vacuum and thermal fluctuations of: (i) the dipole moments of the particles in the chain $p_j^{\rm fl,\pm}(\omega)$ and (ii)  the electric field $E^{\rm fl, \pm}_j(\omega)$. We can write $p^{\pm}_{i}(\omega)$ and $E^{\pm}_{i}(\omega)$ in terms of $p_j^{\rm fl,\pm}(\omega)$ and $E^{\rm fl, \pm}_j(\omega)$ as 
\begin{align}
p_i^\pm(\omega) & = p_{i}^{\rm fl,\pm}(\omega) + \alpha^{\pm}_{{\rm eff},i}(\omega)E_{i}^{\rm fl,\pm}(\omega)+\alpha^{\pm}_{{\rm eff},i}(\omega)\sum^N_{j\neq i}G_{ij}(\omega)p_{j}^{\pm}(\omega) \nonumber, \\
E_i^\pm(\omega) & = E^{\rm fl,\pm}_{i}(\omega) +\sum^N_{j\neq i}G_{ij}(\omega)p_{j}^{\pm}(\omega) +G_0(\omega) p_i^\pm(\omega), \nonumber
\end{align}
where $G_{ij}(\omega)=(1-\delta_{ij})\exp(ikd_{ij})[(kd_{ij})^2+ikd_{ij}-1]/d^3_{ij}$ is the dipole-dipole interaction between particles $i$ and $j$, $k=\omega/c$, and $G_0(\omega) = \frac{2i}{3} k^3$. Notice that the inclusion of $G_0(\omega)$ is necessary to account for radiative corrections \cite{N15}.
Furthermore, $\alpha^{\pm}_{{\rm eff}, i}(\omega)$ is the effective polarizability of particle $i$ as seen from the frame at rest. Following the notation of \cite{N14}, these equations can be solved simultaneously as
\begin{align}
p_i^\pm(\omega) &= \sum_{j=1}^N A_{ij}^\pm(\omega) p_{j}^{\rm fl,\pm}(\omega) + \sum_{j=1}^N B_{ij}^\pm E_{j}^{\rm fl,\pm}(\omega), \nonumber \\
E_i^\pm(\omega) &= \sum_{j=1}^N C_{ij}^\pm(\omega) p_{j}^{\rm fl,\pm}(\omega) + \sum_{j=1}^N D_{ij}^\pm(\omega) E_{j}^{\rm fl,\pm}(\omega), \nonumber
\end{align}
where $A_{ij}^\pm(\omega)$ are the matrix elements of the $N\times N$ matrix defined as $\mathbf{A}^{\pm}(\omega)=\left[\mathcal{I} - \bm{\alpha}^\pm(\omega) \mathbf{G}(\omega)\right]^{-1}$, with $\bm{\alpha}^{\pm}$ being a diagonal matrix whose components are  $\alpha^{\pm}_{{\rm eff},i}(\omega)$, and $\mathbf{G}$ the dipole-dipole interaction matrix with components $G_{ij}(\omega)$.  Similarly, the other matrices are defined as $B_{ij}^\pm(\omega)=A_{ij}^\pm(\omega)\alpha^{\pm}_{{\rm eff},j}(\omega)$, $C_{ij}^\pm(\omega)=\sum_{k=1}^{N}\left[\delta_{ik}G_0(\omega)+G_{ik}(\omega)\right]A_{kj}^\pm(\omega)$, and $D_{ij}^\pm(\omega)=\delta_{ij} +C_{ij}^\pm(\omega)\alpha^{\pm}_{{\rm eff},j}(\omega)$.
Using these expressions, we can write Eq.~(\ref{Miz}) in terms of averages over the dipole and field fluctuations
\begin{equation}\label{Mintpm}
M_{i}^\pm = i\int_{-\infty}^\infty \! \frac{d\omega d\omega'}{4\pi^2} e^{-i(\omega-\omega')t} \sum_{j,k=1}^N \left[ {A_{ij}^\pm}^\ast(\omega')C_{ik}^{\pm}(\omega)\langle {p_{j}^{\rm fl,\pm}}^*(\omega'){p}^{\rm fl,\pm}_{k}(\omega)\rangle
+{B_{ij}^\pm}^*(\omega')D_{ik}^{\pm}(\omega)\langle {E_{j}^{\rm fl,\pm}}^*(\omega')E_{k}^{\rm fl,\pm}(\omega)\rangle\right].
\end{equation}
Notice that there are not cross terms involving dipole and field fluctuations since these are uncorrelated \cite{ama7}. In order to evaluate the averages over fluctuations, we use the fluctuation-dissipation theorem \cite{N1928,CW1951,ama7}, which, for dipole fluctuations, taking into account the rotation of the particles, reads
\begin{equation}\label{FDTP}
\langle {p^{\rm fl,\pm}_{j}}^*(\omega') p^{\rm fl,\pm}_{k}(\omega)\rangle = 4\pi\hbar \delta(\omega-\omega') \chi^{\pm}_{j}(\omega)\delta_{jk} \left[n_{j}(\omega_j^{\mp}) + \frac{1}{2}\right],
\end{equation}
where $\omega_i^\pm = \omega\pm\Omega_i$ and we use $\chi^{\pm}_i(\omega)={\rm Im}\left\{\alpha^{\pm}_{{\rm eff},i}(\omega)\right\}-\frac{2\omega^3}{3c^3}\left|\alpha^{\pm}_{{\rm eff},i}(\omega)\right|^2$ instead of ${\rm Im}\left\{\alpha^{\pm}_{{\rm eff},i}(\omega)\right\}$ to account for the radiative corrections in the response of the nanoparticle \cite{ama18,MTB13}. Similarly, for the field fluctuations, 
\begin{equation}\label{FDTE}
\langle {E^{\rm fl,\pm}_{j}}^*(\omega') E^{\rm fl,\pm}_{k}(\omega)\rangle = 4\pi\hbar \delta(\omega-\omega'){\rm Im} \left\{ \delta_{jk}G_0(\omega)+G_{jk}(\omega)\right\} \left[n_0(\omega)+\frac1{2}\right].
\end{equation}
In these expressions, $n_i(\omega) = \left[\exp(\hbar\omega /k_BT_i)-1\right]^{-1}$ is the Bose-Einstein distribution at temperature $T_i$, while $T_0$ is the temperature of the surrounding vacuum.

With these tools, we can evaluate the averaged dipole and field fluctuations occurring in Eq.~(\ref{Mintpm}). We begin with the first term in that equation, which involves the dipole fluctuations. After using the fluctuation-dissipation theorem given in Eq.~(\ref{FDTP}), we get 
\begin{equation}
M_{i,p}^{\pm} = \frac{i\hbar}{\pi}\int_{-\infty}^\infty \!\! d\omega \sum_{j=1}^N  {A_{ij}^\pm}^*(\omega)C_{ij}^\pm(\omega)\chi^{\pm}_j( \omega) \left[n_j( \omega_{j}^{\mp})+\frac1{2}\right].
\nonumber
\end{equation}
We can simplify the integral over frequency by noting that $n_i(-\omega\pm\Omega_i)+\frac1{2}=-n_i(\omega\mp\Omega_i)-\frac1{2}$ and that, due to causality, $G_{ij}(-\omega)=G_{ij}^*(\omega)$ and $\alpha^{\pm}_{{\rm eff},i}(-\omega) = \alpha^{\mp*}_{{\rm eff},i}(\omega)$, which also imply that $T_{ij}^\pm(-\omega) = {T_{ij}^\mp}^*(\omega)$,  where $T=A$, $B$, $C$, or $D$. By doing so, we obtain 
\begin{equation}
M_{i,p}^{\pm} = -\frac{2\hbar}{\pi}\int_0^\infty \!\! d\omega \sum_{j=1}^N \left[ \frac{2k^3}{3}\left |A_{ij}^\pm(\omega)\right|^2  + \sum_{k=1}^N {\rm Im}\left\{ {A_{ij}^\pm}^*(\omega)A_{ki}^{\pm}(\omega)G_{jk}(\omega)\right\} \right]\chi^{\pm}_j( \omega)\left[n_j( \omega_{j}^{\mp})+\frac1{2}\right].
\nonumber
\end{equation}
Using $\sum_{m=1}^N A_{im}^\pm(\omega)\alpha^{\pm}_{{\rm eff},m}(\omega)G_{mj}(\omega) = A_{ij}^{\pm}(\omega) - \delta_{ij}$, the expression above reduces to
\begin{align}\label{Mzp}
M_{i,p}^{\pm} ={}& -\frac{2\hbar}{\pi}\int_0^\infty \!\! d\omega \left[\frac{2k^3}{3}{\rm Re}\left\{2A^{\pm}_{ii}(\omega)-1 \right\} + {\rm Im}\left\{S^{\pm}_{ii}(\omega)\right\}\right]\chi^{\pm}_{i}(\omega)\left[n_i( \omega_{i}^{\mp})+\frac1{2}\right] \nonumber \\
&+\frac{2\hbar}{\pi}\int_0^\infty \!\! d\omega \sum_{j=1}^N \left|S^{\pm}_{ij}(\omega) \right|^2\chi^{\pm}_{i}(\omega)\chi_{j}^{\pm}(\omega)\left[n_j( \omega_{j}^{\mp})+\frac1{2}\right],  
\end{align}
where $S^{\pm}_{ij}(\omega)=\sum^N_{m=1} A_{mi}^{\pm}(\omega)G_{mj}(\omega)$.

The second term in Eq.~(\ref{Mintpm}) can be computed in a similar way using, in this case, Eq.~(\ref{FDTE}). This leads to
\begin{equation}
M_{i,E}^{\pm} = \frac{i\hbar}{\pi}\int_{-\infty}^\infty \!\! d\omega \sum_{j,k=1}^N {B_{ij}^\pm}^*(\omega)D_{ik}^{\pm}(\omega){\rm Im}\left\{ G_0(\omega) \delta_{jk} + G_{jk}(\omega)\right\} \left[n_0( \omega)+\frac1{2}\right].
\nonumber
\end{equation}
Following the same steps as above and using the definitions of $B_{ij}^\pm(\omega)$ and $D_{ij}^\pm(\omega)$, this expression reduces to 
\begin{align}\label{MzE}
M_{i,E}^{\pm} ={}& \frac{2\hbar}{\pi}\int_0^\infty \!\! d\omega \left[\frac{2k^3}{3}{\rm Re}\left\{2A^{\pm}_{ii}(\omega)-1 \right\} + {\rm Im}\left\{S^{\pm}_{ii}(\omega)\right\}\right]\chi^{\pm}_{i}(\omega)\left[n_0( \omega)+\frac1{2}\right] \nonumber \\
&-\frac{2\hbar}{\pi}\int_0^\infty \!\! d\omega \sum_{j=1}^N \left|S^{\pm}_{ij}(\omega) \right|^2\chi^{\pm}_{i}(\omega)\chi^{\pm}_{j}(\omega)\left[n_0( \omega)+\frac1{2}\right]. 
\end{align}
Finally, combining Eqs.~(\ref{Mzp}) and (\ref{MzE}), we obtain the expression of the Casimir torque given in the paper.\\

\subsection{Polarizability of a Rotating Particle in the Rest Frame}

The polarizability of a nonrotating nanosphere, arising from an optical resonance, can be modeled using a harmonic oscillator model. Within that approximation, the motion of the charges that give rise to the resonance obeys the following equation of motion \cite{ama37}
\begin{equation}
\mathbf{\ddot{r}}(t)=-\omega_0^2 \mathbf{r}(t)-\gamma \mathbf{\dot{r}}(t)+\frac{2Q^2}{3Mc^3}\mathbf{\dddot{r}}(t)+\frac{Q}{M} \mathbf{E}(t). \nonumber
\end{equation}
Here, $\omega_0$ is the frequency of the optical resonance, $\gamma$ is the nonradiative damping of the resonance, $Q$ and $M$ are the total charge and mass of the oscillating charges, and $\mathbf{E}$ is the external field. The third term on the right-hand side corresponds to the Abraham-Lorentz force and describes the radiative damping. If the particle is set to rotate around the $z$ axis with an angular velocity $\mathbf{\Omega} = \Omega \mathbf{\hat{z}}$, the equation of motion above becomes 
\begin{equation}
\mathbf{\ddot{r}}(t)=-\omega_0^2 \mathbf{r}(t) +\Omega^2  [\mathbf{r}(t)-\mathbf{\hat{z}}(\mathbf{\hat{z}}\cdot\mathbf{r}(t))]-\gamma \mathbf{\dot{r}}(t)+\frac{2Q^2}{3Mc^3}\mathbf{\dddot{r}}(t)+\frac{Q}{M} \mathbf{E}(t), \label{eqmrot}
\end{equation}
where the extra term arises from the centripetal acceleration. This equation of motion can be transformed to the frame rotating with the nanoparticle. By doing so, and using $\sim$ to denote the variables in the rotating frame, we obtain
\begin{align}
\mathbf{\ddot{\tilde{r}}}(t)={}&-2\mathbf{\Omega}\times \mathbf{\dot{\tilde{r}}}(t)- \mathbf{\Omega}\times\left(\mathbf{\Omega}\times \mathbf{\tilde{r}}(t)\right)-\omega_0^2\mathbf{\tilde{r}}(t)+\Omega^2  [\mathbf{\tilde{r}}(t)-\mathbf{\hat{z}}(\mathbf{\hat{z}}\cdot\mathbf{\tilde{r}}(t))]-\gamma\left(\mathbf{\dot{\tilde{r}}}(t) +\mathbf{\Omega}\times \mathbf{\tilde{r}}(t) \right) +\frac{Q}{M} \mathbf{\tilde{E}}(t) \nonumber \\
&+\frac{2Q^2}{3Mc^3}\left[\mathbf{\dddot{\tilde{r}}}(t)+3\mathbf{\Omega}\times \mathbf{\ddot{\tilde{r}}}(t)+3\mathbf{\Omega}\times\left(\mathbf{\Omega}\times \mathbf{\dot{\tilde{r}}}(t)\right)-\Omega^2\mathbf{\Omega}\times \mathbf{\tilde{r}}(t)\right], \nonumber
\end{align}
where the new terms on the right-hand side are associated with the Coriolis force and the centrifugal forces. Changing to the circular basis  defined as: $\mathbf{\hat{e}}_{\pm} = (1/\sqrt{2})(\mathbf{\hat{x}}\pm i\mathbf{\hat{y}})$, and $\mathbf{\hat{e}}_0=\mathbf{\hat{z}}$, the equation of motion for the $\pm$ components becomes
\begin{align}
\ddot{\tilde{r}}^{\pm}(t) ={}&\pm 2i\Omega\dot{\tilde{r}}^{\pm}(t)+\Omega^2\tilde{r}^{\pm}(t) -(\omega_0^2-\Omega^2) \tilde{r}^{\pm}(t)-\gamma \dot{\tilde{r}}^{\pm}(t)\pm i\gamma\Omega \tilde{r}^{\pm}(t)+\frac{Q}{M} \tilde{E}^{\pm}(t) \nonumber  \\
&+\frac{2Q^2}{3Mc^3}\left[\mp 3i\Omega \ddot{\tilde{r}}^{\pm}(t) -3\Omega^2\dot{\tilde{r}}^{\pm}(t) \pm i \Omega^3 \tilde{r}^{\pm}(t) \right]. \nonumber
\end{align}
Assuming that $\tilde{E}^{\pm}(t)$, and therefore $\tilde{r}^{\pm}(t)$, oscillate at frequency $\omega$, we can calculate the $\pm$ components of the polarizability of the nanoparticle in the rotating frame
\begin{equation}
\tilde{\alpha}^{\pm}(\omega)=\frac{\frac{3c^3}{2\omega_0^2}\gamma_r}{\omega_0^2-\Omega^2-(\omega\pm\Omega)^2-i\gamma(\omega\pm\Omega)-i\gamma_r\frac{(\omega\pm\Omega)^3}{\omega_0^2}}, \nonumber
\end{equation}
where we have defined $\gamma_r=(2Q^2\omega_0^2)/(3Mc^3)$. We can transform this polarizability to the frame at rest to obtain the following effective polarizability
\begin{equation}
\alpha_{\rm eff}^{\pm}(\omega)=\frac{\frac{3c^3}{2\omega_0^2}\gamma_r}{\omega_0^2-\Omega^2-\omega^2-i\gamma\omega-i\gamma_r\frac{\omega^3}{\omega_0^2}}. \nonumber
\end{equation}
Notice that this  equation can also be directly obtained from Eq.~(\ref{eqmrot}). In summary, the effect of rotation is a shift of the resonance, which for $\Omega\ll \omega_0$ becomes $\omega_0-\Omega^2/(2\omega_0)$.



\subsection{Analysis of the Thermal Equilibrium Approximation}

Throughout the manuscript, we assume that the nanoparticles remain at the same temperature as the environment. To verify the validity of this approximation, we can analyze the rate of change of the rotational energy of the nanoparticles, $P_{\rm rot}$, which can be written as $P_{\rm rot}= \sum^N_{i=1}\Omega_{i} M_{i}\leq0$ in terms of the angular velocity and the torque acting on the particles. The energy lost from the decrease in rotation has to be either radiated or converted into an increase of the temperature of the nanoparticles. The full calculation of the coupled thermal and mechanical dynamics of the particles is beyond the scope of this work. However, we can estimate the maximum increase in the temperature of the nanoparticles from $P_{\rm rot}+P_{\rm rad} (T_0+\Delta T)=0$, where $P_{\rm rad} (T_0+\Delta T)$ is the power radiated by the nanoparticles to the environment when their temperatures are $\Delta T$ above that of the environment, $T_0$. 
If this equality is satisfied, all the energy lost from the decrease in rotation is radiated to the environment, and, therefore, the temperature of the particles does not further increase. It is worth noting that the value of $\Delta T$ obtained from this equality is an overestimation, since it does not account for the energy required to reach that temperature.

\begin{figure}[H]
\begin{center}
\includegraphics[width=175mm,angle=0,clip]{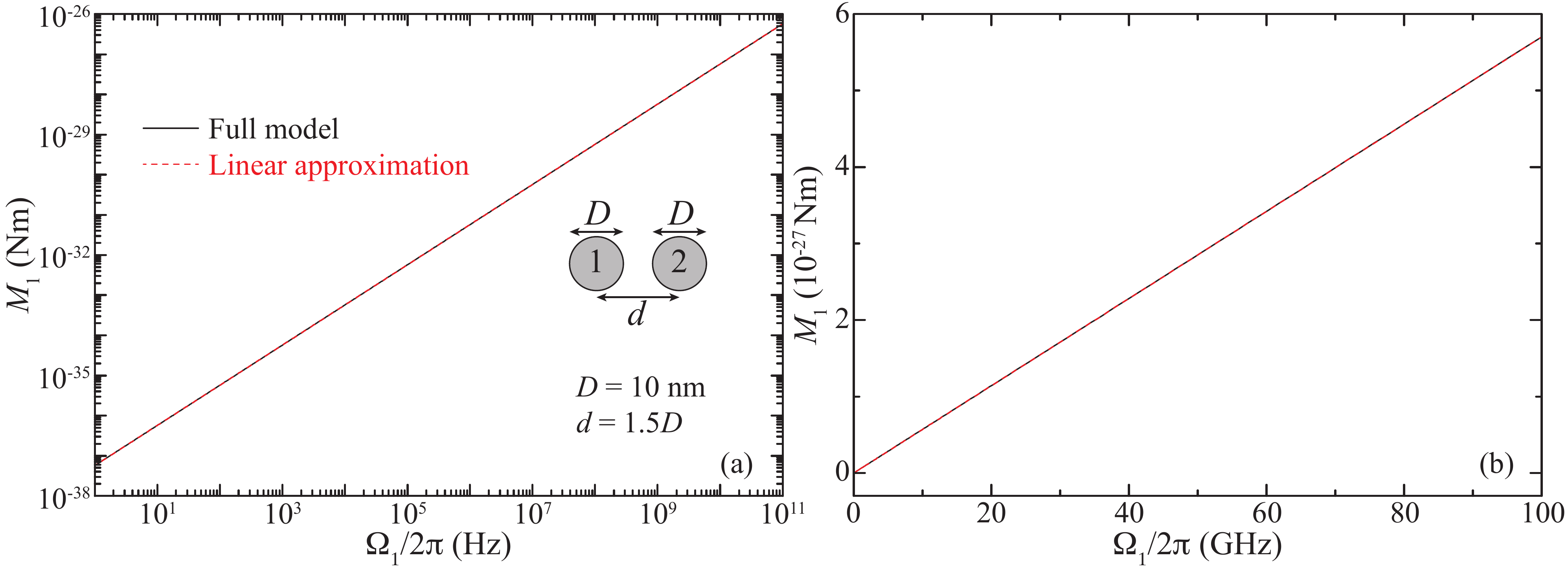}
\caption{ Validity of linear approximation in the calculation of the torque. (a) Torque, plotted as a function of the angular velocity,  for a rotating SiC nanoparticle of $D=10\,$nm placed at a distance $d=1.5D$ from an identical nanoparticle at rest. The solid black curve represents the results obtained using the full model ignoring terms $\mathcal{O}(\Omega^2)$ in the effective polarizability, while the dashed red curve corresponds to the results of the linear approximation. (b) Zoom of the results shown in (a) over the GHz region. In both panels the agreement between the two approaches is perfect.} \label{figS1}
\end{center}
\end{figure}

\begin{figure}[H]
\begin{center}
\includegraphics[width=175mm,angle=0,clip]{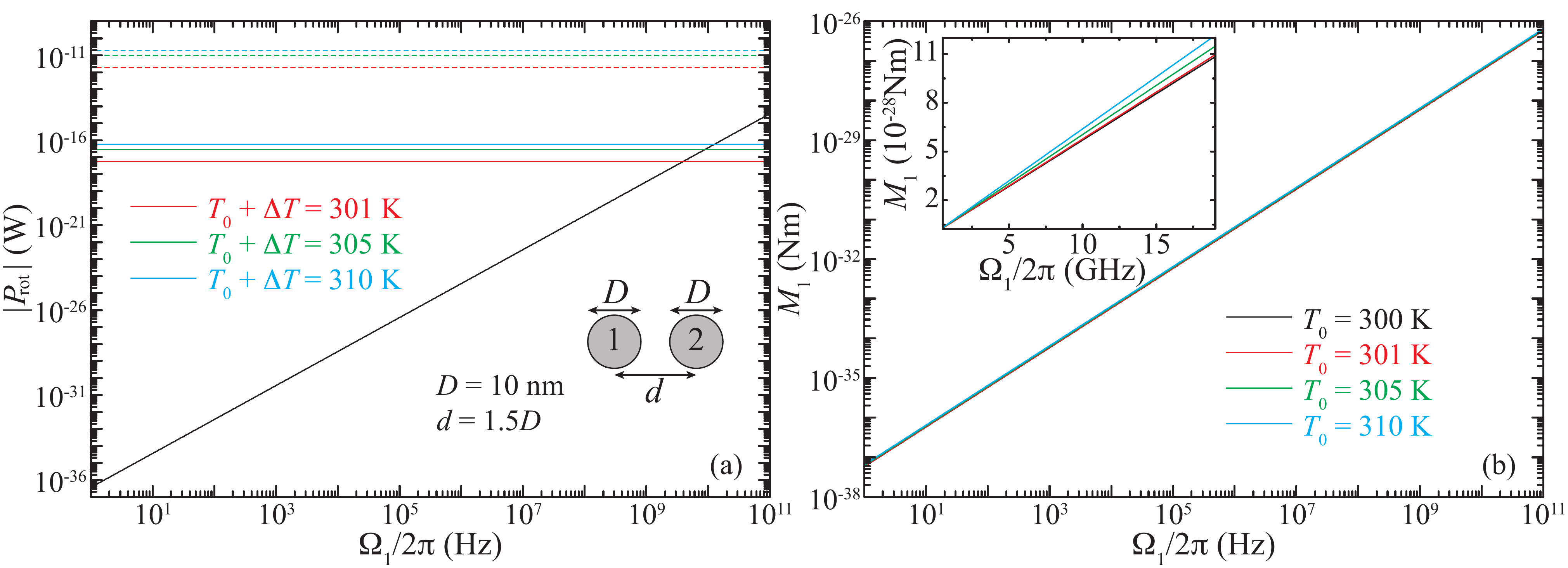}
\caption{Thermal effects on the Casimir torque. (a) The black curve shows the absolute value of the rate of change of the rotational energy, $|P_{\rm rot}|$, for a rotating SiC nanoparticle of $D=10\,$nm placed at a distance $d=1.5D$ from an identical nanoparticle at rest. We assume that the particles and the environment are at a temperature of $300\,$K. The solid red, green, and blue curves represent the power radiated by the two nanoparticles to the environment when they are at temperature $T_0+\Delta T=301\,$K, $305\,$K, and $310\,$K,  respectively,  while the environment is at $T_0=300\,$ K. The dashed red, green, and blue curves represent the power transferred from particle $1$ to particle $2$ when the former is at $301\,$K, $305\,$K, or $310\,$K, while the latter and the environment are kept at $300\,$K. (b) Torque for the system considered in (a), calculated assuming the particles and the environment are at temperature $T_0=300\,$K (black curve), $301\,$K (red curve), $305\,$K (green curve), or $310\,$K (blue curve). The inset shows a zoom of the results over the GHz region. All of the results are obtained with the full model ignoring terms of $\mathcal{O}(\Omega^2)$ in the effective polarizability.} \label{figS2}
\end{center}
\end{figure}

The black solid line in Fig.~\ref{figS2}(a) shows $P_{\rm rot}$ for a rotating SiC nanoparticle of diameter $D=10\,$nm placed at a distance $d=1.5D$ from an identical nanoparticle at rest. This calculation, which is performed using the full model neglecting terms $\mathcal{O}(\Omega^2)$ in the effective polarizability (see  Fig.~\ref{figS1}), assumes the particles and the environment to be at $300\,$K. 
The value of $P_{\rm rot}$ is compared with the power radiated by the two nanoparticles, $P_{\rm rad}(T_0+\Delta T)$, when $T_0+\Delta T=301\,$K, $305\,$K, and $310\,$K. The corresponding results are indicated by the solid red, green, and blue curves. We calculate $P_{\rm rad}(T_0+\Delta T)$ following the approach of \cite{ama18}, and, therefore, we do not consider the effect of the rotation, which, for $\Omega_i\ll k_B T_0/\hbar$, is expected to be negligible.
Examining these results, we observe that $\Delta T$ is always smaller than $5\,$K for the angular velocities under consideration. This change in temperature does not appreciably alter the value of the torque, as shown in panel (b), where we calculate the torque on particle $1$ assuming that both particles and the environment are at temperature $300\,$K (black curve), $301\,$K (red curve), $305\,$K (green curve), or $310\,$K (blue curve).  

It is important to notice that this approach assumes that all of the particles have the same temperature, which is, indeed, a very good approximation since the transfer of energy between the nanoparticles happens several orders of magnitude faster than the transfer of energy from them to the environment. This can be seen by looking at the dashed curves on Fig.~\ref{figS2}(a), which signal the rate of energy transfer from particle $1$ to particle $2$ when the former is at temperature $T_0+\Delta T$ and the latter is at the same temperature as the environment, $T_0$. 


\subsection{Degenerate Decay Rates}

The natural modes of the system considered in Figure~4(c) of the main paper are 
\begin{equation}
\mathbf{\hat{e}}_1 = \begin{pmatrix} 0.577 \\ 0.577 \\ 0.577 \end{pmatrix}
\qquad
\mathbf{\hat{e}}_2 = \begin{pmatrix} 0.707 \\ 0.605\times 10^{-14} \\ -0.707 \end{pmatrix}
\qquad
\mathbf{\hat{e}}_3 = \begin{pmatrix} 0.706 \\ -0.055 \\ 0.706 \end{pmatrix},\nonumber
\end{equation}
where we are using a notation in which the first, second, and third components correspond, respectively, to the angular velocity of the first, second, and third particles. The associated decay rates are $\lambda_1 = -0.331\times 10^{-6}\, {\rm s}^{-1}$ and $\lambda_2 = \lambda_3 = -0.085\,$s$^{-1}$.
If we take a superposition of modes $2$ and $3$, we can build a rotation state that decays with the same rate $\lambda_2=\lambda_3$. Consider, for instance, the following initial angular velocity for the chain: $\mathbf{\Omega}(0) = c_2\mathbf{\hat{e}}_2+c_3\mathbf{\hat{e}}_3$, where $c_3=1.001 c_2$. With this superposition, we have that $\Omega_3(0) = 0$, while  $\Omega_1(0) = 1.414c_2$ and  $\Omega_2(0) = -0.055c_2$. Therefore, this choice of initial angular velocities forces particle $3$ to be at rest for the entire course of the dynamics. We can alternatively choose an initial condition such that particle $3$ is forced to rotate at any desired angular frequency $\Omega_c$. This can be achieved by including a component proportional to $\mathbf{\hat{e}}_1$, \textit{i.e.}, $\mathbf{\Omega}(0) = c_1\mathbf{\hat{e}}_1+ c_2\mathbf{\hat{e}}_2+c_3\mathbf{\hat{e}}_3$.
By doing so, we find $\Omega_3(0) = 0.577c_1=\Omega_c$, while $\Omega_1(0) = 1.414c_2+0.577c_1$ and  $\Omega_2(0) = - 0.055c_2+0.577c_1$.

\clearpage

\begin{figure}
\begin{center}
\includegraphics[width=175mm,angle=0,clip]{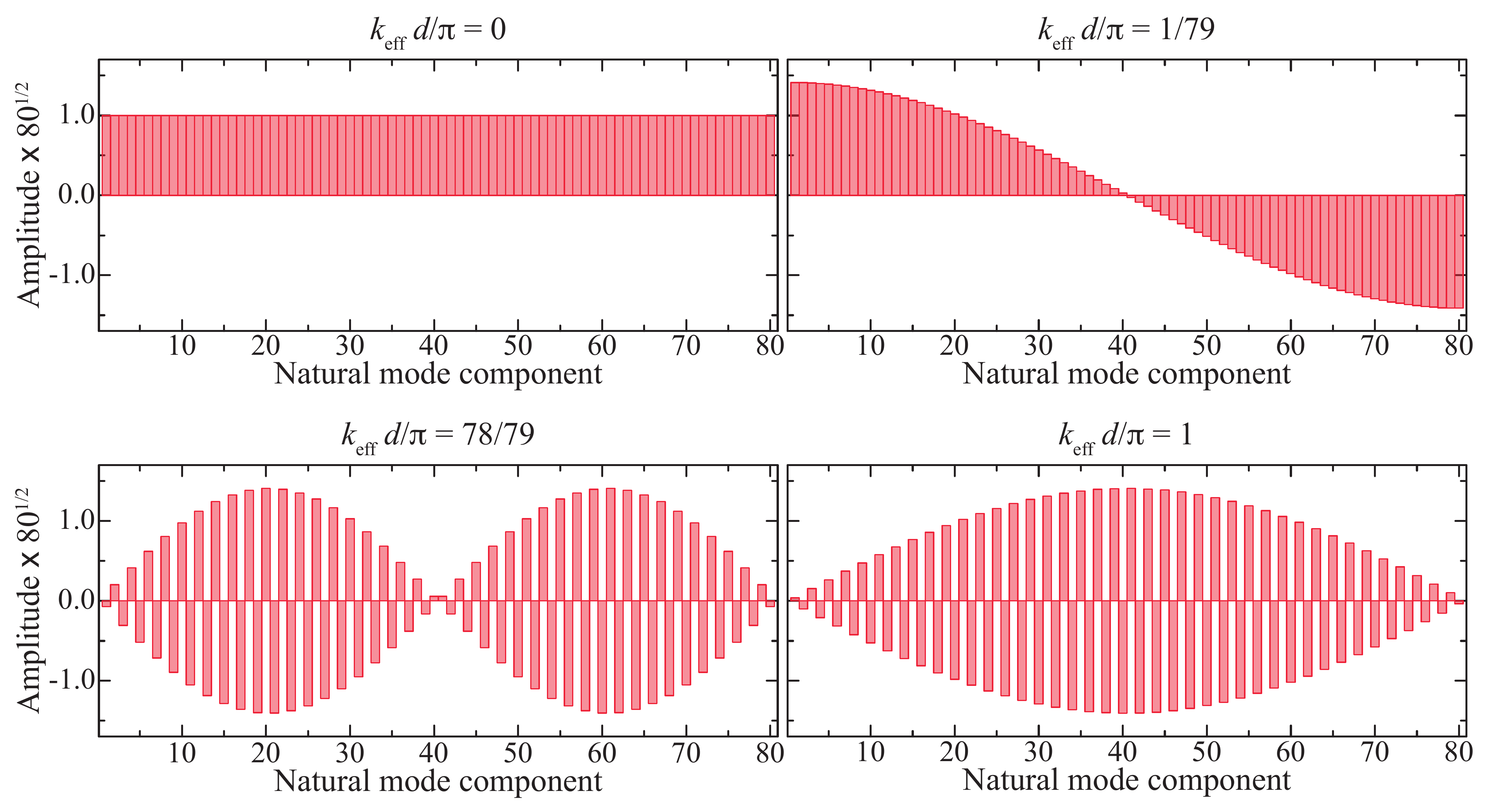}
\caption{ Amplitude of four different natural modes of the longest chain studied in Figure~3 of the main paper.  The chain contains $N=80$ SiC particles with $D=10\,$nm and $d=1.5D$. The upper row shows the first and second modes, while, in the lower row, the next-to-last and the last modes are displayed. The corresponding effective momentum is indicated above each plot.} \label{figS3}
\end{center}
\end{figure}

\begin{figure}
\begin{center}
\includegraphics[width=175mm,angle=0,clip]{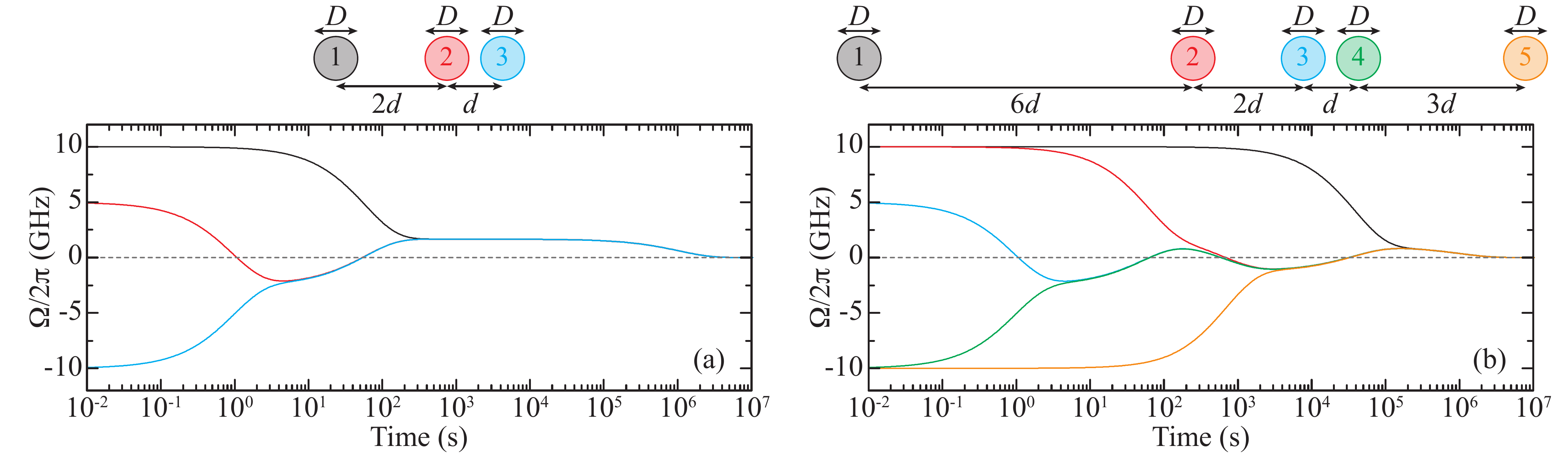}
\caption{Exotic rotational dynamics. (a,b) Temporal evolution of the angular frequency of the particles of a chain with $N=3$ (a) and $N=5$ (b), arranged as shown in the upper schematics. In both cases, $D=10\,$nm and $d=1.5D$.} \label{figS4}
\end{center}
\end{figure}

\begin{figure}
\begin{center}
\includegraphics[width=140mm,angle=0,clip]{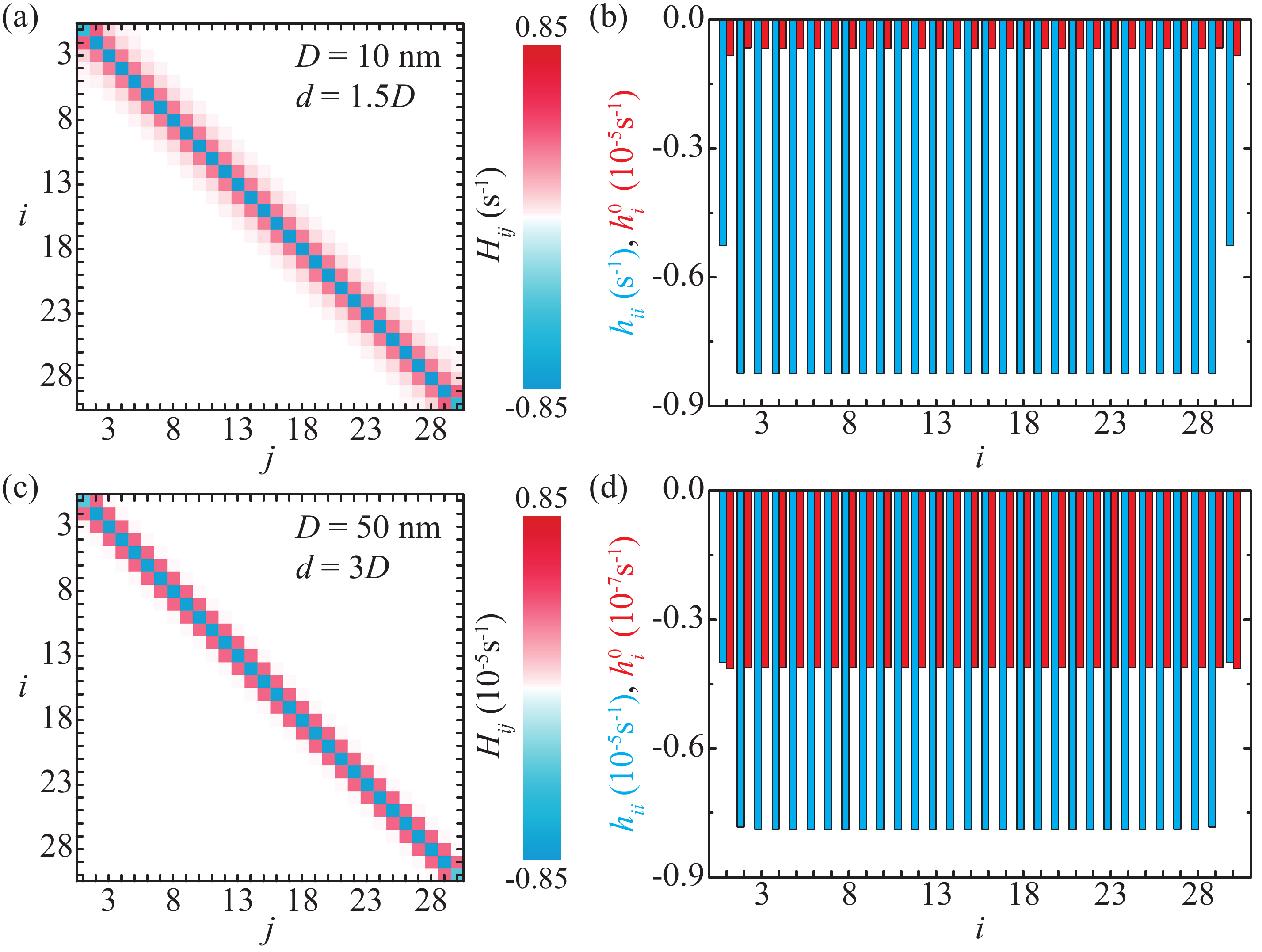}
\caption{Rotational dynamics. (a) $H_{ij}$ for a chain of $N=30$ identical particles with $D=10\,$nm made of SiC, which are uniformly distributed with a center-to-center distance $d=1.5D$. (b) Diagonal components of $h_{ij}$ (blue) and the corresponding $h^0_{i}$ (red). (c,d) Same as (a,b), but for a chain with $D=50\,$nm and $d=3D$. Notice the different scales in the plots.} \label{figS5}
\end{center}
\end{figure}

\clearpage

%

\end{document}